\documentclass[twoside]{ilcws10}
\usepackage[latin1]{inputenc}
\usepackage[dvips]{graphicx,epsfig,color}
\usepackage{wrapfig,rotating}
\usepackage{amssymb,amsmath,array}
\usepackage{url}

\begin{document}
\title{
Shintake Monitor in ATF2: Performance Evaluation} 
\author{Yohei Yamaguchi$^1$ Takashi Yamanaka$^1$ Masahiro Oroku$^1$ Yoshio Kamiya$^2$ Sachio Komamiya$^{1, 2}$\\ Taikan Suehara$^2$ Toshiyuki Okugi$^3$ Nobuhiro Terunuma$^3$ Toshiaki Tauchi$^4$ Sakae Araki$^3$ Junji Urakawa$^3$
\vspace{.3cm}\\
1- The University of Tokyo - Department of Physics \\
7-3-1 Hongo, Bunkyo-ku, Tokyo 113-0033 - Japan
\vspace{.1cm}\\
2- The University of Tokyo - ICEPP \\
7-3-1 Hongo, Bunkyo-ku, Tokyo 113-0033 - Japan
\vspace{.1cm}\\
3- KEK - Accelerator Laboratry\\
1-1 Oho, Tsukuba, Ibaraki 305-0801 - Japan
\vspace{.1cm}\\
4- KEK - Institute of Particle and Nuclear Science\\
1-1 Oho, Tsukuba, Ibaraki 305-0801 - Japan\\
}

\maketitle

\begin{abstract}
The beam test for the Shintake monitor succeeded in measuring signal modulation with the laser interference fringe pattern in November 2009.
We have studied the error sources, and evaluated the systematic error to be less than 30\% for 1 minute measurements.
This paper centers on the evaluation of the Shintake monitor performance through analyzing beam tests deta.
Most systematic error sources are well understood, enabling accurate measurement of the beam size when it reaches 37 nm.
\end{abstract}

\section{ATF2}
ATF2 is the the final focus test facility for the ILC to realize and demonstrate nanometer focusing based on a local chromaticity collection~\cite{ATF}.
The first goal of ATF2 is focusing the beam to nanometer scale (37 nm in design) in vertical.
The second goal is to stabilise the beam focal point at the few-nanometre level for a long period.
In order to measure such a small beam, Shintake monitor has been developed for ATF2 beam.

\section{Shintake Monitor}

\begin{wrapfigure}{r}{0.5\columnwidth}
\centerline{\includegraphics[width=0.45\columnwidth ]{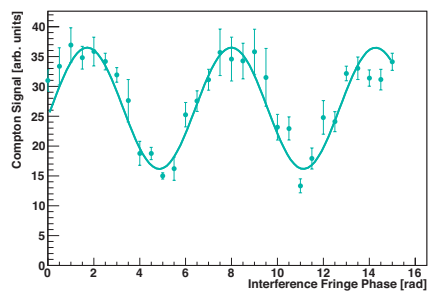}}
\caption{Beam size measurement}
\label{Fig:fringe_scan}
\end{wrapfigure}

The Shintake monitor is a beam size monitor which employs a fringe pattern from laser interference~\cite{Shintake}.

When the fringe pattern scans electron beam, the number of scattered photons $N_{\gamma}$ is modulated in the following manner
\begin{equation}
N_{\gamma} \propto 1+M\cos\left(2k_{y}y+\alpha\right) .
\label{eq:principle}
\end{equation}
Here, $\alpha$ is the phase of the interference fringe, $k_y$ is vertical component of the wave number and $M$ is the modulation depth which represents modulation magnitude.
Figure \ref{Fig:fringe_scan} shows the fringe scan at beam test in ATF2.

$M$ relates to vertical beam size as the following:
\begin{equation}
M=\left|\cos\theta\right|\exp\left(-2k_{y}^{2}\sigma_{y}^{2}\right) ,
\label{eq:modulation}
\end{equation}
where $\sigma_y$ is the vertical beam size and $\theta$ is the laser crossing angle.
Since the modulation depth is obtained by the measurement, the beam size can be calculated from Equation (\ref{eq:modulation}).

In ATF2, we scan the beam by changing the phase of the interference fringe pattern step by step.
This method enables us to measure the beam size without moving the laser paths nor the electron beam orbit~\cite{Yamanaka}.

\section{Beam Tests}
\begin{wrapfigure}{r}{0.5\columnwidth}
\centerline{\includegraphics[width=0.45\columnwidth ]{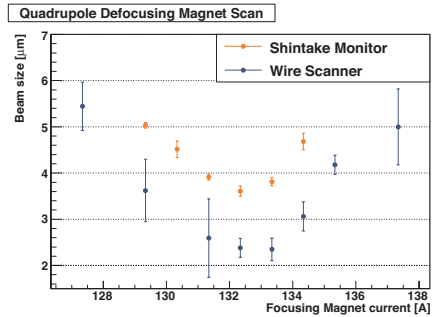}}
\caption{Comparison with the wire scanner measurement}
\label{Fig:comp_mwip}
\end{wrapfigure}

The Shintake monitor was installed at IP of ATF2, and the commissioning of ATF2 started in the beginning of 2009.
In November 2009 we found signal modulation with the laser interference fringe pattern by the Shintake monitor.
In order to evaluate the accuracy of beam size measurement by the Shintake monitor, we compared the Shintake monitor beam size measurement with the wire scanner (10 $\mu$m$\phi$) measurement.
Figure \ref{Fig:comp_mwip} shows that the Shintake monitor measurement is always larger than that by wire scanner.
Systematic error seems to exist for the Shintake monitor, beacause the wire scanner measurement is reliable above 2.5 $\mu$m.

\section{Systematic Error}
\begin{wraptable}{l}{0.5\columnwidth}
\centerline{\begin{tabular}{|l|c|c|}
\hline
$C_i$ & by 2009 & future \\
      &         & (for 37 nm) \\ \hline
$C_{pol}$ & 96.1$\pm$2.1\%  & 99.7$\pm$0.1\% \\
$C_{align}$ & $>$97.5\%  & --- \\
$C_{phase}$  & 92.3$\pm$ 2.0\%  & 95.0$\pm$1.2\% \\
$C_{temp}$ & $>$99.7\%  & --- \\
$C_{spherical}$ & 100\%  & $>$99.7\% \\
$C_{growth}$ & 100\%  & $\sim$98.5\% \\ \hline
$\Pi_i  C_{i}$ & $>$86.3\%  & $>$90.4\% \\
\hline
\end{tabular}}
\caption{Modulation reduction factors}
\label{tab:C_i}
\end{wraptable}

Several sources of systematic error reduce modulation depth. Their influence is expressed as
\begin{equation}
M_{meas.}=C_{\alpha}C_{\beta}\cdots M_{ideal},
\end{equation}
where $M_{meas.}$ is measured modulation depth, $M_{ideal}$ is ideal modulation depth and $C_i$s are modulation reduction factors.
The measured modulation depth is shown as the product of all modulation reduction factors and the ideal modulation depth.

Table \ref{tab:C_i} shows evaluation of $C_i$s.

\subsection{Laser Polarization}
In principle, laser polarization never reduces contrast of the interference fringe.
However, because the laser beam splitter is tuned for s-polarized light, p-polarized reflectance of the splitter is not exactly 50\%.
Thus the existence of p-polarized light causes laser power imbalance and reduces the contrast of the fringe pattern.
This contrast degradation in turn reduces signal modulation.
The modulation reduction factor from polarization is written as
\begin{equation}
C_{pol}=\frac{2\left(\sqrt{P_{1s}P_{2s}}+\sqrt{P_{1p}P_{2p}}\right)}{P},
\end{equation}
where $P_s$ is s-polarized laser power, $P_p$ is p-polarized laser power, $P$ is the combined power of the laser and the number of subscripts shows the light path.
During the beam test in the December 2009, the modulation reduction factor was estimated to be 96.3\%.
After the beam test, we adjusted laser polarization with half wave plate.
Currently the factor is nearly 100\%.

\subsection{Laser Alignment Accuracy}
In this section, we adopt an approximation in which the laser size in horizontal direction is much larger than the horizontal beam size.
Furthermore, we assume that the laser beam profile takes on a Gaussian distribution.

\subsubsection{Longitudinal}
If two laser lights differ in spot size, laser power imbalance occurs locally.
We denote the modulation reduction factor coming from longitudinal profile imbalance as $C_{z,profile}$.
They are written as the following:
\begin{equation}
C_{z,profile} = \sqrt{\frac{2\sigma_{1z,laser}\sigma_{2z,laser}}{\sigma_{1z,laser}^{2}+\sigma_{2z,laser}^{2}}} ,
\label{eq:C-zpro}
\end{equation}
where $\sigma_{z,laser}$ is the laser spot size in longutusinal direction.

If two lasers overlap only partially, the modulation becomes small.
The modulation reduction factor from misalignment of pathway is written as
\begin{equation}
C_{z,path}=
\exp \left(-\frac{\Delta z^{2}}{8 \sigma_{z,laser}^{2} }\right) \\
\label{eq:C-zpath}
\end{equation}
where $\Delta z$ is spatial difference between two lasers in longitudinal direction.

\subsubsection{Transverse}
In the same way as Equation (\ref{eq:C-zpro}) and (\ref{eq:C-zpath}), We can evaluate the modulation reduction factors from laser misalignment on transverse plane, $C_{t,profile}$ and $C_{t,path}$.
However, they are not independent from each other.
Therefore if two laser lights are different in both spot sizes and position, the modulation reduction factor cannot be shown as simply the product of $C_{t,profile}$ by $C_{t,path}$ but as
\begin{equation}
C_{t,align}=\frac{2\sqrt{\sigma_{1t,laser}\sigma_{2t,laser}}}{\sigma_{2t,laser}\exp\left[-\left(\frac{l_{1}}{2\sigma_{1t,laser}}\right)^{2}\right]+\sigma_{1t,laser}\exp\left[\left(\frac{l_{1}}{2\sigma_{1t,laser}}\right)^{2}\right]} .
\end{equation}
where $\sigma_{t,laser}$ is the transverse laser size and $\Delta t$ is the spatial difference between two lasers on transverse plane.

During beam tests, we align the laser pathway and profile with electron beam position by analyzing beam size measurement results.
We demand the alignment accuracy to have a modulation reduction factor exceeding 97.5\%.

\subsection{Phase Jitter}
Because the signal energy jitter caused by the relative position jitter between the electron beam and the laser interference fringe pattern is biased towards the mean value of the signal energy, the modulation depth is systematically reduced by the relative position jitter.
The modulation reduction factor is written as
\begin{equation}
C_{phase}= \exp\left(-\frac{\Delta \alpha^{2}}{2}\right)\cos\left(2k_{y}y+\alpha\right) ,
\end{equation}
where $\Delta \alpha$ is phase jitter converted from a relative position jitter.
Phase jitter is induced by optical devices vibrations.
When optical devices vibrate, each laser path length shifts slightly.
Because differences in laser path length determine the fringe phase, this vibration causes phase jitter.

The measured phase jitter was about 400 mrad.
The Shintake monitor has a phase monitor and feedback system of light path length with piezo drive.
Using the feedback system, phase jitter was reduced to 320 mrad, after which the contrast reduction factor was recalculated to be 95\%.

\subsection{Laser Temporal Coherence}
If the laser temporal coherence is poor and the two laser path lengths are different, the contrast of the interference fringe is reduced, and is the signal modulation.
This is because each laser frequency component contributes to interference in different phases under the existence of laser path length difference.
The modulation reduction factor due to this effect is written as
\begin{equation}
C_{temp} = \exp\left(-2\pi^{2}\left(\frac{\delta\nu\Delta l}{c}\right)^{2}\right) ,
\end{equation}
where $\frac{\delta \nu}{c}$ is line width of the laser and $\Delta l$ is laser path length difference.
The line width of the laser we use is less than 0.003 cm$^{-1}$.
We evaluate the modulation reduction factor from laser temporal coherence to be larger than 99.7\%.

\subsection{Laser Spherical Wavefront}
Gaussian beams have spherical wavefronts.
Wavefront curvature radius decreases with distance from focal point.
Therefore if the electron beam position is away from the laser focal point when passing the fringe pattern, beam senses curved fringe pattern due to the laser spherical wavefront.
When laser rays cross at 180 degrees, this effect is given by the following expression:
\begin{equation}
C_{spherical}=\left(1+\Delta\hat{y}^{2}\right)^{-\frac{1}{4}}\left[1+\Delta\hat{y}^{2}\left(1+z_{R}\frac{1+\Delta\hat{y}^{2}}{2k\sigma_{x}^{2}}\right)^{-2}\right]^{-\frac{1}{4}} ,
\end{equation}
where $z_R$ is Rayleigh length between beam and laser focal point in vertical axis, $\Delta \hat{y}$ is a distance normalized by the Rayleigh length and $k$ is wave number of a laser light.
To reduce this effect, we need to align laser focal point.

\subsection{Beam Size Growth in the Fringe Pattern}
Due to small beta function of the electron beam under strong focusing, beam size growth within the fringe pattern is significant.
The modulation reduction factor is written as
\begin{equation}
C_{growth}=\left(1+4k_{y}^{2}\sigma_{z}^{2}\frac{\epsilon_y}{\beta^{\star}_y}\right)^{-\frac{1}{2}} ,
\end{equation}
where $\epsilon_y$ is vertical emittance and $\beta^{\star}_y$ is beta function at IP.
To evaluate $C_{growth}$, beam optics study is necessary at both upper and lower IP.

\subsection{Tilt of the Laser Fringe Pattern}
When the horizontal or longitudinal axes of the interference fringe pattern are not parallel to the electron beam axes, the measured beam size ends up larger than the actual value.
In the current beam condition, this effect is the most significant one.
The measured beam size is written as
\begin{equation}
\sigma_{y}^{2} \rightarrow \sigma_{y}^{\prime2}
= \sigma_{y}^{2} \cos^{2} \Delta \varphi_{t}
+ \frac{\sigma_{x}^{2} \sin^{2} \Delta \varphi_{t} }{1+\sigma_{x}^{2}\sigma_{t,laser}^{-2}\sin^{2}\phi}
\end{equation}
\begin{equation}
\sigma_{y}^{2} \rightarrow \sigma_{y}^{\prime2}
= \sigma_{y}^{2} \cos^{2} \Delta \varphi_{z}
+ \sigma_{z,laser}^{2} \sin^{2} \Delta \varphi_{z} ,
\end{equation}
where $\sigma_y^{\prime}$ is measured beam size, $\sigma_y$ is an ideal beam size, $\delta \varphi_z$ is angle difference between the fringe longitudinal axis and the beam dirrection, $\sigma_x$ is horizontal beam size and $\phi$ is half laser cross angle.

During 2009 beam test, the electron beam angle was tuned for the wire scanner.
So if the fringe pattern and the wire were not parallel to each other, there should exist systematic error between the two measurement results.
We estimate that the two measurements would be consistent, if the angle difference between the fringe and the wire is about 2 degrees.

To reduce this influence, we need to align the laser light path angles.
Because the laser crossing point must not be changed by this alignment, we have to align some mirror angles delicately.

\section{Moving Towards the Ultimate 37nm Beam Size Measurement}
Through estimation of all these error sources, we evaluate the Shintake monitor performance towards 37 nm beam size measurement which is one of the ATF2 goals.
From our calculation, systematic error is estimated to be about 3\%.
This satisfies the demand of ATF2.
We shall strive to produce a 37 nm vertical waist and measure the size after the summer shutdown.


\begin{footnotesize}


\end{footnotesize}



\begin{thebibliography}{99}
\bibitem{slide}Slide\\
\url{http://ilcagenda.linearcollider.org/contributionDisplay.py?contribId=59&sessionId=6&confId=4175}
\bibitem{ATF}ATF2 Collaboration, ATF2 Proposal, KEK Report 2005-2 (2005)
\bibitem{Shintake}T. Shintake, Nucl. Inst. and Meth. {\bf A311} (1992) 453.
\bibitem{Yamanaka}T. Yamanaka, LCWS/ILC2010\\
\url{http://ilcagenda.linearcollider.org/contributionDisplay.py?contribId=58&sessionId=6&confId=4175}
\bibitem{Shintake2}P. Tenenbaum and T. Shintake, Annu. Rev. Nucl. Part. Sci. 49, 125 (1999)
\end{thebibliography}
\end{document}